\pdfoutput=1
\documentclass[11pt]{article}

\usepackage[a4paper,margin=25mm]{geometry}
\usepackage[T1]{fontenc}
\usepackage[utf8]{inputenc}
\usepackage{textcomp}
\usepackage{amsmath,amssymb,bm,mathrsfs}
\usepackage{graphicx}
\usepackage[font=small,labelfont=bf]{caption}
\usepackage[numbers,sort&compress]{natbib}
\usepackage[colorlinks=true,linkcolor=blue,citecolor=blue,urlcolor=blue]{hyperref}
\usepackage{microtype}

\graphicspath{{figures/}}
\title{Superconductivity, pseudogap and marginal Fermi liquid in a relative-momentum-local theory}
\author{Jian-Jian Miao\\[0.5em]
\normalsize Quantum Science Center of Guangdong-Hong Kong-Macao Greater Bay Area,\\ \normalsize Shenzhen 518045, China\\[0.35em]
\normalsize\href{mailto:miaojianjian@quantumsc.cn}{\nolinkurl{miaojianjian@quantumsc.cn}}}
\date{}

\begin{document}
\maketitle

\begin{abstract}
Cuprate superconductors host \(d\)-wave superconductivity (SC), a pseudogap (PG) with Fermi arcs and a strange metal (SM) with incoherent excitations and temperature-linear resistivity. These regimes emerge from the same doped Mott system but are usually described separately. Here a relative-momentum-local (RML) theory is constructed, with the exact spin-singlet Cooper representation of antiferromagnetic superexchange in the \(t\)-\(J\) model serving as a common thread. Pair correlations organize the low-energy behaviour of this Cooper channel into phase-coherent static SC, phase-incoherent quasi-static PG and phase-incoherent dynamic SM. A static RML prescription yields an exactly solvable PG Hamiltonian with nodal electron poles, antinodal gaps, resolution-broadened Fermi arcs and gapless charge-\(2e\) Cooper surfaces. A Cooper-surface Lifshitz transition generates logarithmic enhancements in the specific-heat coefficient and charge response without a spin singularity. The particle-particle continuum at finite centre-of-mass momentum produces Ohmic damping, while a dynamic RML prescription yields a one-loop self-energy with marginal-Fermi-liquid scaling. Spectral-weight transfer in the physical pair spectrum diagnoses the PG--SM crossover, whereas nonzero phase stiffness and global phase coherence identify SC. The RML theory relates single-particle, pair-sensitive and thermodynamic probes to three infrared organizations of one Cooper channel. A decisive experimental signature would be zero-energy Cooper-surface ridges in two-electron angle-resolved photoemission spectroscopy (2e-ARPES).
\end{abstract}

\section{Introduction}\label{introduction}

Cuprate superconductors pose a stringent theoretical problem: superconductivity (SC), the pseudogap (PG) and the strange metal (SM) emerge from the same doped Mott system, yet display sharply different low-energy organizations.\cite{ref01,ref02} In the SC state, phase-sensitive Josephson interferometry and half-flux quantization in tricrystal geometries establish the sign reversal of a predominantly \(d_{x^2-y^2}\) order parameter.\cite{ref03,ref04,ref05} This result fixes a non-negotiable symmetry benchmark: a microscopic framework must account not only for a \(d\)-wave pairing amplitude, but also for phase stiffness and global coherence.\cite{ref06,ref07,ref08} In the underdoped normal state, angle-resolved photoemission spectroscopy (ARPES) shows that low-energy electron spectral weight survives first near the zone diagonals, whereas the antinodes remain gapped and the apparent Fermi surface is truncated into arcs rather than closed contours.\cite{ref09,ref10,ref11,ref12,ref13} The arcs lengthen with temperature.\cite{ref14} Their visibility also depends on linewidth, matrix elements and experimental resolution, and substantial arc weight can therefore be non-quasiparticle in character.\cite{ref09,ref15} Across compounds and doping levels, the distinct evolution of nodal and antinodal spectral energies, weights and linewidths, together with the separation of PG and SC gap scales, defines a nodal--antinodal dichotomy that any common theoretical description must reproduce.\cite{ref16,ref17,ref18} Near optimal doping, the SM exhibits \(\rho_{ab}(T)=\rho_0+A_1T\) over an unusually broad window, with linearity extending to low temperature when SC is suppressed.\cite{ref02,ref19,ref20,ref21,ref22} Optical spectroscopy reveals a broad non-Drude response and anomalous frequency-dependent relaxation.\cite{ref23,ref24} In the superconducting state of underdoped cuprates, symmetry-resolved \(B_{1g}\) and \(B_{2g}\) Raman spectra reveal distinct antinodal and nodal energy scales and quasiparticle dynamics.\cite{ref25} ARPES likewise finds broad line shapes and approximately \(T\)- and \(|\omega|\)-linear single-particle scattering near optimal doping, departing from Fermi-liquid behaviour over the measured normal-state window.\cite{ref09,ref26,ref27} These observables are complementary rather than interchangeable: a linear single-particle decay rate does not by itself establish \(T\)-linear dc transport, which also requires momentum relaxation and the appropriate current vertex.\cite{ref28}

The natural theoretical starting point is a doped Mott insulator. Anderson' s resonating-valence-bond (RVB) proposal identified short-range spin singlets as the parent structure from which mobile carriers and superconductivity emerge, while the single-band \(t\)-\(J\) model provides a minimal effective Hamiltonian for this picture.\cite{ref29,ref30,ref31,ref32} Gutzwiller-renormalized mean-field theory yields a projected \(d\)-wave BCS saddle, and early slave-boson treatments obtained a superexchange-driven \(d\)-wave state while distinguishing spinon pairing from charge coherence.\cite{ref33,ref34,ref35,ref36} Spin-fluctuation pairing provides a complementary route to a \(d\)-wave state.\cite{ref37,ref38,ref39} The separation between a large pairing scale and a smaller stiffness scale motivates phase-fluctuation descriptions of the underdoped regime.\cite{ref40} Related BCS--BEC crossover and pairing-fluctuation theories describe incoherent-pair regimes.\cite{ref41} The PG and SM have nevertheless been described within largely separate theoretical frameworks. Phase-disordered pairing, finite-momentum pairing, competing or intertwined orders, gauge theories of fractionalization and phase-string theory provide distinct routes to the PG.\cite{ref32,ref40,ref42,ref43,ref44,ref45,ref46,ref47,ref48,ref49,ref50} In the phenomenological Yang--Rice--Zhang (YRZ) theory of the PG, the coherent part of the Green' s function is \(G_{\mathrm{YRZ}}(\mathbf k,\omega)=g_t(x)[\omega-\xi_{\mathbf k}-\Delta_R^2(\mathbf k)/(\omega+\xi^0_{\mathbf k})]^{-1}\), with zero-frequency zeros on \(\xi^0_{\mathbf k}=0\) away from the \(d\)-wave nodes and spectrally asymmetric pole pockets that appear as Fermi arcs.\cite{ref51} The SM is often summarized by the marginal-Fermi-liquid (MFL) ansatz \(\Sigma_{\mathrm{MFL}}^R(\omega,T)\simeq\lambda[\omega\ln(X/\Omega_c)-i(\pi/2)X]\), with \(X=\max(|\omega|,k_{\mathrm B}T)\),\cite{ref52} whereas antiferromagnetic (AFM) quantum criticality, formulated through spin-fermion theories of electrons coupled to critical spin-density-wave fluctuations, offers another route to non-Fermi-liquid dynamics.\cite{ref53,ref54,ref55} Despite their individual successes, a common description of SC, PG and MFL phenomenology as distinct infrared organizations of one Cooper channel remains lacking.

Here a \emph{relative-momentum-local} (RML) theory is constructed, with the exact spin-singlet Cooper representation of the AFM superexchange interaction in the \(t\)-\(J\) model serving as a common thread. The no-double-occupancy constraint is treated analytically in the Gutzwiller approximation. The block pair correlator \(\mathcal P_{\mathbf q\mathbf q'}(\mathbf Q,i\Omega_m)\), the collective pair correlator \(\mathcal C_{\alpha\alpha'}(\mathbf Q,i\Omega_m)\), and the physical pair spectrum \(\mathcal A^{\mathrm{pair}}_{\alpha\alpha}(\mathbf Q,\Omega)\) resolve three infrared organizations of the same Cooper channel. SC is phase-coherent and static, with a zero-centre-of-mass (COM) momentum \(d\)-wave singularity and nonzero phase stiffness. The PG is phase-incoherent and quasi-static: a static RML prescription yields the exactly solvable block Hamiltonian \(\hat H_{\mathrm{PG}}\) with nodal electron poles, antinodal gaps, resolution-broadened Fermi arcs and gapless charge-\(2e\) Cooper surfaces. The SM is phase-incoherent and dynamic: transverse intersections of the particle-particle continuum at finite COM momentum generate Ohmic damping in the Gaussian approximation, while a dynamic RML prescription yields the minimal effective action \(S_s\), whose one-loop fermion self-energy has marginal-Fermi-liquid scaling. Spectral-weight transfer within \(\mathcal A^{\mathrm{pair}}\) provides an operational diagnostic of the PG--SM crossover, whereas phase stiffness distinguishes the SC transitions. The RML theory relates single-particle, pair-sensitive and thermodynamic probes to three infrared organizations of one Cooper channel. The predicted zero-energy Cooper-surface ridges would provide a direct target for two-electron angle-resolved photoemission spectroscopy (2e-ARPES).

\protect\phantomsection\label{three-infrared-organizations}
\section{One Cooper channel, three organizations}\label{one-cooper-channel-three-organizations}

The RML theory takes the AFM superexchange interaction as its starting point, with the exact spin-singlet Cooper representation serving as the common thread for the three infrared organizations. The central exact operator identity is
\protect\phantomsection\label{eq-main-1}
\begin{equation}
\label{eq:1}
\hat H_J
        =J\sum_{\langle ij\rangle}
        \left(\hat{\mathbf S}_i\!\cdot\!\hat{\mathbf S}_j
        -\frac14\hat n_i\hat n_j\right)
        =-\frac{2J}{N_s}\sum_{\mathbf Q,\alpha=s,d}
        \sum_{\mathbf q,\mathbf q'}
        \phi_{\mathbf q,\alpha}\phi_{\mathbf q',\alpha}
        \hat\Delta_{\mathbf Q,\mathbf q}^{\dagger}
        \hat\Delta_{\mathbf Q,\mathbf q'}.
\end{equation}
Here \(\hat{\mathbf S}_i=\frac12\sum_{\sigma\sigma'}\hat c_{i\sigma}^{\dagger}\boldsymbol\tau_{\sigma\sigma'}\hat c_{i\sigma'}\) and \(\hat n_i=\sum_\sigma\hat c_{i\sigma}^{\dagger}\hat c_{i\sigma}\) are the electron spin and number operators on site \(i\), respectively; \(\hat c_{i\sigma}\) is the electron annihilation operator on site \(i\) with spin \(\sigma\), and \(\boldsymbol\tau\) denotes the vector of Pauli matrices. \(\langle ij\rangle\) runs over nearest-neighbour bonds, \(J>0\) is the AFM superexchange coupling, and \(N_s\) is the number of lattice sites. \(\hat\Delta_{\mathbf Q,\mathbf q}=(\hat c_{\mathbf Q/2+\mathbf q,\uparrow}\hat c_{\mathbf Q/2-\mathbf q,\downarrow}-\hat c_{\mathbf Q/2+\mathbf q,\downarrow}\hat c_{\mathbf Q/2-\mathbf q,\uparrow})/\sqrt2\) is the spin-singlet Cooper-pair operator with COM momentum \(\mathbf Q\) and relative momentum \(\mathbf q\), and \(\phi_{\mathbf q,\alpha}=(\cos q_x+\eta_\alpha\cos q_y)/2\) is the square-lattice form factor, with \(\eta_s=+1\) and \(\eta_d=-1\) for the extended-\(s\)- and \(d\)-wave channels, respectively. In the pairing representation, the AFM superexchange describes dynamical Cooper-pair fluctuations as coherent scattering that conserves COM momentum: an incoming spin-singlet Cooper pair \((\mathbf Q,\mathbf q')\) is scattered into an outgoing one \((\mathbf Q,\mathbf q)\). Rewriting the AFM superexchange in terms of the collective-pair operator \(\hat\Lambda_{\mathbf Q,\alpha}=N_s^{-1}\sum_{\mathbf q}\phi_{\mathbf q,\alpha}\hat\Delta_{\mathbf Q,\mathbf q}\), the interaction \(\hat H_J=-2JN_s\sum_{\mathbf Q,\alpha}\hat\Lambda_{\mathbf Q,\alpha}^{\dagger}\hat\Lambda_{\mathbf Q,\alpha}\) is attractive in each \((\mathbf Q,\alpha)\) sector, thereby favouring pairing in the infrared.

Within the RML theory, SC, PG and SM are distinguished through the Cooper-pair correlation function
\protect\phantomsection\label{eq-main-2}
\begin{equation}
\label{eq:2}
\mathcal P_{\mathbf q\mathbf q'}(\mathbf Q,i\Omega_m)
        =
        \int_0^\beta d\tau\,
        e^{i\Omega_m\tau}
        \left\langle
        T_\tau
        \hat\Delta_{\mathbf Q,\mathbf q}^{\dagger}(\tau)
        \hat\Delta_{\mathbf Q,\mathbf q'}(0)
        \right\rangle,
\end{equation}
where \(\Omega_m=2\pi m/\beta\) is the bosonic Matsubara frequency, \(\beta=1/T\) with \(k_{\mathrm B}=1\) throughout, and \(T_\tau\) denotes imaginary-time ordering. The block-diagonal component \(\mathcal P_{\mathbf q\mathbf q}\) measures pair autocorrelation within the individual relative-momentum block \(\{\mathbf q,-\mathbf q\}\), whereas an off-diagonal component \(\mathcal P_{\mathbf q\mathbf q'}\), with \(\mathbf q\ne\pm\mathbf q'+\mathbf G\) for every reciprocal-lattice vector \(\mathbf G\), measures inter-block pair correlation between distinct blocks. However, nonzero inter-block correlations alone do not imply SC. The macroscopic SC coherence is instead diagnosed by the collective-pair correlation function
\protect\phantomsection\label{eq-main-3}
\begin{equation}
\label{eq:3}
\begin{aligned}
        \mathcal C_{\alpha\alpha'}
        (\mathbf Q,i\Omega_m)
        &=
        \frac{1}{N_s}
        \sum_{\mathbf q,\mathbf q'}
        \phi_{\mathbf q,\alpha}\,
        \mathcal P_{\mathbf q\mathbf q'}
        (\mathbf Q,i\Omega_m)\,
        \phi_{\mathbf q',\alpha'}
        \\
        &=
        N_s\int_0^\beta d\tau\,
        e^{i\Omega_m\tau}
        \left\langle
        T_\tau
        \hat\Lambda_{\mathbf Q,\alpha}^{\dagger}(\tau)
        \hat\Lambda_{\mathbf Q,\alpha'}(0)
        \right\rangle.
        \end{aligned}
\end{equation}
An SC phase in the \(\alpha\)-channel with long-range order (LRO) gives \(\mathcal C_{\alpha\alpha}(\mathbf 0,0)=O(N_s)\), whereas a normal state with finite correlation length and time gives only \(O(1)\). In two dimensions (2D), the Berezinskii--Kosterlitz--Thouless (BKT) phase instead has algebraically decaying phase coherence and nonzero renormalized stiffness, but no LRO. The SC organization is characterized as phase-coherent and static in the \(d\)-wave channel: \(\mathcal C_{dd}(\mathbf Q,i\Omega_m)\) develops a coherent \(\delta\)-function contribution at \((\mathbf Q,i\Omega_m)=(0,0)\) for LRO, or an algebraic infrared singularity in the BKT phase, accompanied by a macroscopic collective mode and nonzero phase stiffness. The PG organization is characterized as phase-incoherent and quasi-static: the block-diagonal weight \(\mathcal P_{\mathbf q\mathbf q}\) is enhanced, while the collective \(d\)-wave \(\mathcal C_{dd}\) forms a finite-width peak around \((\mathbf Q,i\Omega_m)=(0,0)\), with \(\Delta Q\sim\xi_{\mathrm{PG}}^{-1}\) and \(\Delta\Omega\sim\tau_{\mathrm{PG}}^{-1}\). However, the leading eigenvalue of collective \(\mathcal C_{\alpha\alpha'}\) remains \(O(1)\) for finite Cooper-pair correlation length and time, and the phase stiffness vanishes. The SM organization is characterized as phase-incoherent and dynamic: after analytic continuation, the retarded pair spectral function \(\mathcal A_{\alpha\alpha}^{\mathrm{pair}}(\mathbf Q,\Omega)=-\pi^{-1}\operatorname{Im}\mathcal C_{\alpha\alpha}^{R}(\mathbf Q,\Omega)\) remains broad over a finite range of COM momenta and nonzero real frequencies. The gapless particle-particle continuum supplies Ohmic damping, and the dynamic RML prescription can yield a marginal-Fermi-liquid self-energy with a sufficiently isotropic and approximately temperature-independent RML coupling, while the phase stiffness remains zero. The PG-SM boundary is therefore a spectral-weight transfer crossover, whereas the PG-SC and SM-SC boundaries are phase transitions marked by the onset of nonzero phase stiffness and a coherent static \(d\)-wave infrared singularity.

\begin{figure}[tbp]
\centering
\includegraphics[width=\textwidth,height=0.48\textheight,keepaspectratio]{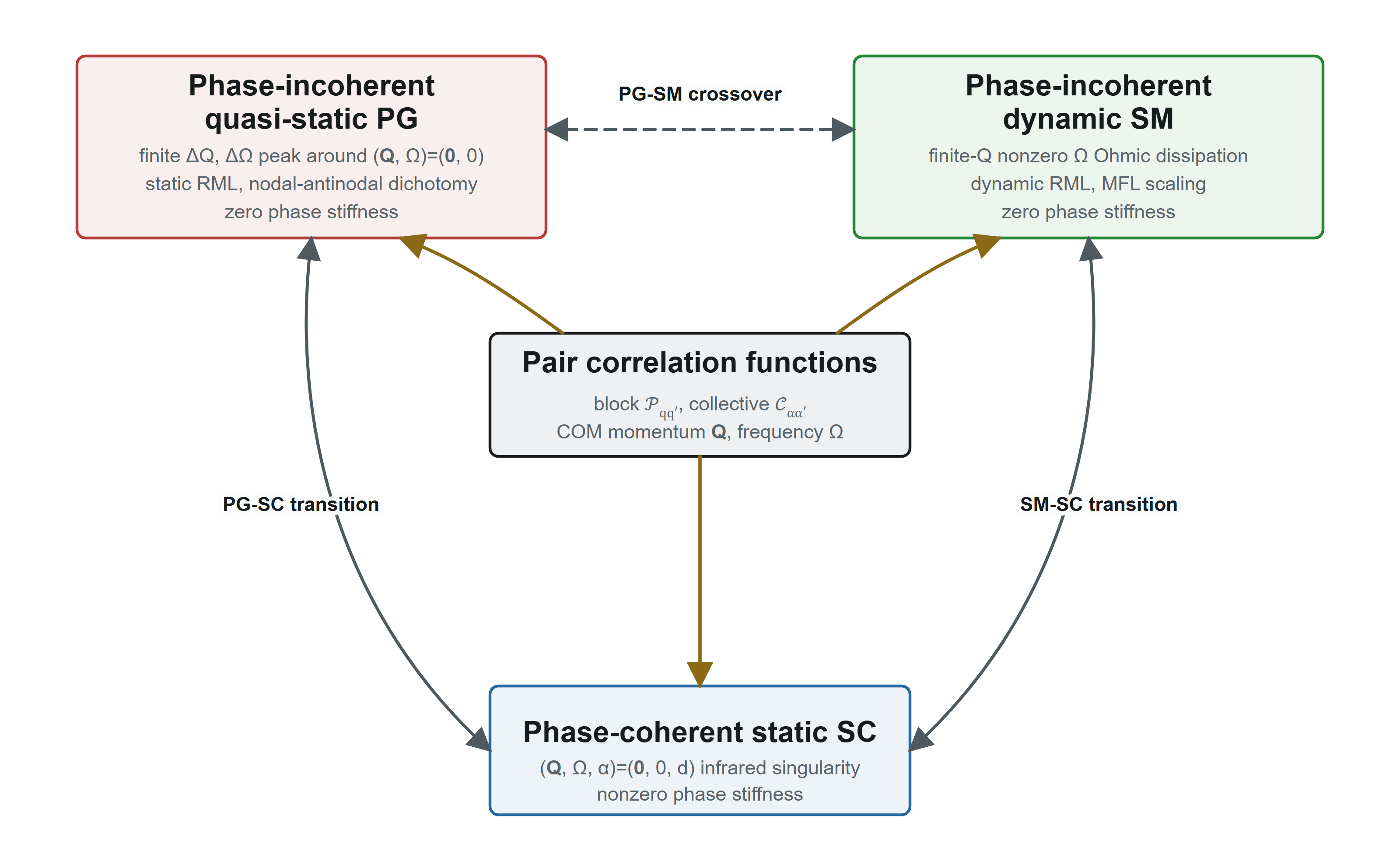}
\caption{\textbf{Three infrared organizations of pair correlation functions.} The block-resolved Cooper-pair correlation \(\mathcal P_{\mathbf q\mathbf q'}\) and the channel-resolved collective-pair correlation \(\mathcal C_{\alpha\alpha'}\) distinguish three infrared organizations. In the PG organization, enhanced block-diagonal pair autocorrelation \(\mathcal P_{\mathbf q\mathbf q}\) produces a quasi-static, finite-width peak in \(\mathcal C_{dd}\) around \((\mathbf Q,i\Omega_m)=(0,0)\), with finite \(\Delta Q\) and \(\Delta\Omega\) and no phase stiffness. In the SM organization, the retarded pair spectral function \(\mathcal A_{\alpha\alpha}^{\mathrm{pair}}(\mathbf Q,\Omega)\) remains broad over finite COM momenta and nonzero real frequencies. The gapless particle-particle continuum supplies Ohmic damping, and the dynamic RML prescription can yield a marginal-Fermi-liquid self-energy with a sufficiently isotropic and approximately temperature-independent RML coupling, while the phase stiffness remains zero. In the SC organization, \(\mathcal C_{dd}\) develops a coherent \(\delta\)-function contribution with \(\mathcal C_{dd}(\mathbf 0,0)=O(N_s)\) for LRO, or an algebraic infrared singularity in the 2D BKT phase, with nonzero phase stiffness. The dashed bidirectional arrow denotes the PG-SM crossover, while the solid bidirectional arrows denote the PG-SC and SM-SC transitions.}
\label{fig:1}
\end{figure}

\protect\phantomsection\label{tj-d-wave-sc}
\section{t-J model and d-wave SC}\label{t-j-model-and-d-wave-sc}

In the plain vanilla RVB picture, the cuprates are viewed as doped Mott insulators whose low-energy physics is governed by carrier motion subject to the no-double-occupancy constraint and by AFM superexchange. The microscopic \(t\)-\(J\) model captures these essential ingredients compactly via a single-band effective Hamiltonian. The analytically intractable Gutzwiller projection enforcing the no-double-occupancy constraint is treated within the Gutzwiller approximation. This yields the renormalized Hamiltonian \(\hat H_{t\text{-}J}^{G}=g_t(x)\hat H_t+g_J(x)\hat H_J\). Here \(g_t(x)=2x/(1+x)\) and \(g_J(x)=4/(1+x)^2\) are the doping-dependent renormalization factors.\cite{ref33} For a time-reversal-invariant normal state with a generic Fermi surface, an arbitrarily weak attraction in the Cooper channel is marginally relevant under the Fermi-surface renormalization group, and the \(\mathbf Q=0\) Cooper sector carries the leading logarithmic infrared divergence.\cite{ref56} Therefore, to obtain the SC organization, only the uniform \(\mathbf Q=0\) sector is retained, and the SC order parameter \(\Phi_\alpha=2g_J(x)J\langle\hat\Lambda_{\mathbf 0,\alpha}\rangle\) is introduced by the mean-field decoupling of \(\hat H_J\) in the \(\alpha\)-channel. The extended-\(s\)- and \(d\)-wave channels belong to distinct irreducible representations of the square-lattice \(D_{4h}\) point group. For the parameters used here and at fixed \(x\), the self-consistent \(d\)-wave saddle has a lower Helmholtz free energy than the extended-\(s\)-wave saddle in the underdoped regime. In 2D, the mean-field pairing scale \(\Phi_\alpha\) and the BKT phase-ordering temperature \(T_{\mathrm{BKT}}\) are generally distinct. Integrating out the massive amplitude fluctuations about the \(d\)-wave saddle yields a short-distance phase-only action with bare stiffness \(\rho_d^{(0)}\). A rough estimate gives \(T_{\mathrm{BKT}}^{(0)}=\frac{\pi}{2}\rho_d^{(0)}(T_{\mathrm{BKT}}^{(0)})\), where the long-distance vortex-renormalized stiffness \(\rho_d^R\) in the Nelson--Kosterlitz jump is replaced by the bare stiffness \(\rho_d^{(0)}\).\cite{ref57} The doping dependences of the mean-field pairing amplitude onset \(T_{\Delta_d}\) and the bare phase-ordering scale \(T_{\mathrm{BKT}}^{(0)}\) are shown in Fig.~\ref{fig:2}a.

\protect\phantomsection\label{pseudogap-fermi-arc}
\section{Pseudogap and Fermi arc}\label{pseudogap-and-fermi-arc}

In quasi-2D cuprates, the unbinding of SC vortices disorders the global phase coherence of the collective mode \(\Phi_\alpha\) near the SC transition temperature, whereas a local pairing amplitude can still remain above \(T_c\).\cite{ref07,ref08,ref40} However, BKT phase disordering does not erase the off-diagonal coherent scattering. A postulated static RML prescription implements this essential inter-block dephasing and block factorization, yielding the exactly solvable PG Hamiltonian:
\protect\phantomsection\label{eq-main-4}
\begin{equation}
\label{eq:4}
\hat H_{\mathrm{PG}}
        =\sum_{\mathbf q}'\left[
        \xi_{\mathbf q}(\hat n_{\mathbf q}+\hat n_{-\mathbf q})
        -2\mathcal J_{\mathbf q}(x,T)
        \hat\Delta_{\mathbf q}^{\dagger}\hat\Delta_{\mathbf q}
        \right],
\end{equation}
where \(\hat n_{\mathbf q}=\sum_\sigma\hat c_{\mathbf q\sigma}^{\dagger}\hat c_{\mathbf q\sigma}\) and \(\hat\Delta_{\mathbf q}=\hat\Delta_{\mathbf Q=0,\mathbf q}\) are the momentum-space number operator and the zero-COM Cooper-pair operator, respectively. The static \(d\)-wave coupling satisfies \(\mathcal J_{\mathbf q}(x,T)=\sqrt{2}\,\Phi_d(x,T)\phi_{\mathbf q,d}^{2}\ge0\), and is strictly positive away from the \(d\)-wave nodal lines. The prime restricts the relative-momentum sum to one representative from each generic \(\{\mathbf q,-\mathbf q\}\) block. \(\hat H_{\mathrm{PG}}\) is understood as an effective description of the pseudogap in the temperature regime \(T_{\mathrm{BKT}}<T<T^*\ll T_F\) and in the infrared limits \(|\mathbf Q|/k_F\to0\) and \(|\Omega|/E_F\to0\). The quasi-static PG organization retains the pairing scale \(\Phi_d\) of the mean-field \(d\)-wave saddle.

\(\hat H_{\mathrm{PG}}\) essentially describes a number-conserving normal state. Distinct blocks of \(\hat H_{\mathrm{PG}}\) commute and each generic block is solved exactly in a 16-dimensional Fock space. The ground states in distinct blocks are
\protect\phantomsection\label{eq-main-5}
\begin{equation}
\label{eq:5}
|\mathrm{GS}_{\mathbf q}\rangle=
        \begin{cases}
        |0\rangle,&N_{\mathbf q}=0,\quad \xi_{\mathbf q}>\mathcal J_{\mathbf q},\\
        \hat\Delta_{\mathbf q}^{\dagger}|0\rangle,
        &N_{\mathbf q}=2,\quad 0< \xi_{\mathbf q}< \mathcal J_{\mathbf q},\\
        (\hat\Delta_{\mathbf q}^{\dagger})^2|0\rangle,
        &N_{\mathbf q}=4,\quad \xi_{\mathbf q}<0,
        \end{cases}
\end{equation}
where \(N_{\mathbf q}\) denotes the eigenvalue of the block number operator \(\hat N_{\mathbf q}=\hat n_{\mathbf q}+\hat n_{-\mathbf q}\). The enhanced diagonal weight \(\langle\hat\Delta_{\mathbf q}^{\dagger}\hat\Delta_{\mathbf q'}\rangle=\delta_{\mathbf q\mathbf q'}\Theta(\mathcal J_{\mathbf q}-\xi_{\mathbf q})\) is a pair autocorrelation and does not imply a condensate amplitude because \(\langle\hat\Delta_{\mathbf q}\rangle=0\). \(\mathcal C_{dd}\) has no macroscopic coherent contribution and the phase stiffness vanishes. The exact solvability of \(\hat H_{\mathrm{PG}}\) is an analytical convenience but can still reflect the main PG physics; a physical PG can retain finite short-range inter-block correlations.

\begin{figure}[p]
\centering
\includegraphics[width=0.77\textwidth]{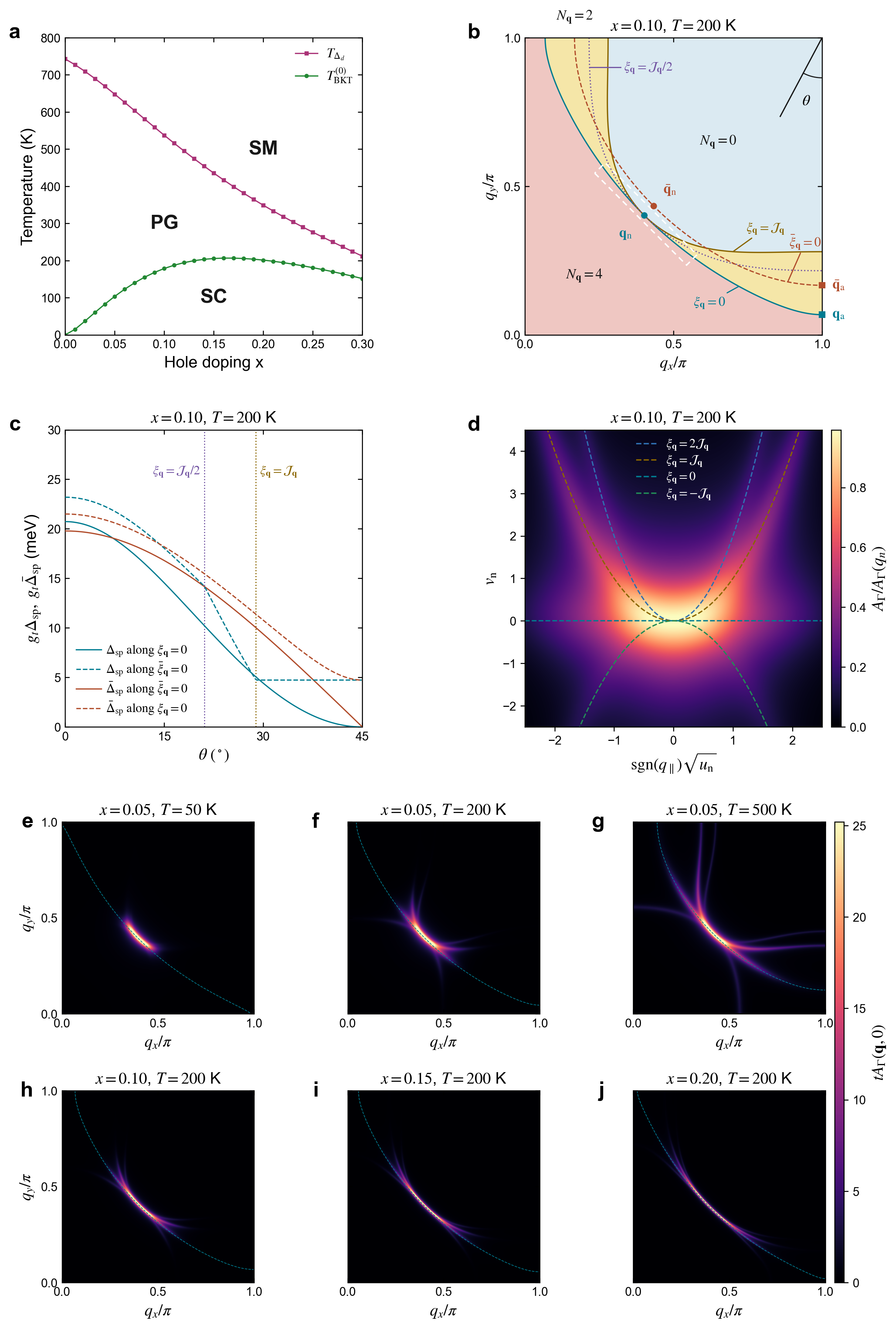}
\captionsetup{width=0.77\textwidth,font=footnotesize,skip=4pt}
\caption{\textbf{BKT, pseudogap, and Fermi arc.} \textbf{a,} Mean-field \(T_{\Delta_d}\), defined as the highest temperature at which \(\hat H_{t\text{-}J}^{G,\mathrm{MF}}\) has a nonzero \(d\)-wave saddle, and bare \(T_{\mathrm{BKT}}^{(0)}\) obtained from the phase-only action with bare phase stiffness, which neglects long-distance vortex renormalization. \textbf{b,} Ground-state occupations of \(\hat H_{\mathrm{PG}}\) at \(x=0.10\). The occupation boundaries are \(\mathcal S_{2e}:\xi_{\mathbf q}=0\) and \(\mathcal S_{2h}:\xi_{\mathbf q}=\mathcal J_{\mathbf q}\), while the punctured Luttinger surface is \(\mathcal S_{\mathrm L}:\xi_{\mathbf q}=\mathcal J_{\mathbf q}/2\). \textbf{c,} The single-particle gap scales \(\Delta_{\mathrm{sp}}\) of the exact PG model and \({\bar\Delta}_{\mathrm{sp}}\) of the mean-field \(d\)-wave SC state, each evaluated along both the PG Fermi contour \(\mathcal S_{\mathrm{PG}}\) and the Fermi surface \(\mathcal S_{\mathrm F}\), with all four curves multiplied by \(g_t(x)=2x/(1+x)\). \textbf{d,} Node-normalized zero-frequency \(A_\Gamma(\mathbf q,0)\) near the PG node at \(x=0.10\) and \(T=200\,\mathrm K\), with dashed guides for the four exact zero-energy Lehmann contours. Here \(u_{\mathrm n}=\kappa_{\mathrm n}q_\parallel^2/\Gamma\) and \(v_{\mathrm n}=v_{F,\mathrm n}q_\perp/\Gamma\). \textbf{e-g,} Zero-energy single-particle spectra at fixed \(x=0.05\) for \(T=50,200,500\,\mathrm K\). \textbf{h-j,} The corresponding doping evolution at \(T=200\,\mathrm K\) for \(x=0.10,0.15,0.20\). The cyan dashed line marks the PG Fermi contour \(\xi_{\mathbf q}(T)=0\) at fixed density \(n=1-x\). The calculations use \(t=0.4\,\mathrm{eV}\), \(t'/t=-0.22\), \(t''/t=0.045\), \(J/t=0.16\), and a \(720\times720\) cell-centred grid; the displayed spectral maps use \(480\times480\) sampling and \(\Gamma/t=0.01\;(4.0\,\mathrm{meV})\).}
\label{fig:2}
\end{figure}

For comparison between the PG and \(d\)-wave SC states, the following notation is adopted: \(\mathcal S_{\mathrm{PG}}\) is the Fermi contour of \(\hat H_{\mathrm{PG}}\) defined by \(\xi_{\mathbf q}=0\), with PG node \(\mathbf q_{\mathrm n}\), antinode \(\mathbf q_{\mathrm a}\), and single-particle gap \(\Delta_{\mathrm{sp}}\); \(\mathcal S_{\mathrm F}\) is the Fermi surface of \(\hat H_{t\text{-}J}^{G,\mathrm{MF}}\) defined by \({\bar\xi}_{\mathbf q}=0\), with node \({\bar{\mathbf q}}_{\mathrm n}\), antinode \({\bar{\mathbf q}}_{\mathrm a}\), and single-particle gap \({\bar\Delta}_{\mathrm{sp}}\) (Fig.~\ref{fig:2}b). The PG single-particle spectrum is gapped everywhere except at \(\mathbf q_{\mathrm n}\), where \(\xi_{\mathbf q_{\mathrm n}}=\mathcal J_{\mathbf q_{\mathrm n}}=0\). Along \(\mathcal S_{\mathrm{PG}}\), \(\Delta_{\mathrm{sp}}\propto\phi_{\mathbf q,d}^{2}\) and is maximal at \(\mathbf q_{\mathrm a}\); by contrast, along \(\mathcal S_{\mathrm F}\), \({\bar\Delta}_{\mathrm{sp}}\propto|\phi_{\mathbf q,d}|\) and is maximal at \({\bar{\mathbf q}}_{\mathrm a}\). The PG and SC chemical potentials \(\mu_{\mathrm{PG}}(x)\) and \(\mu_d(x)\) are both determined at fixed density \(n=1-x\), resulting in the same minimum gap \(\lvert\mu_{\mathrm{PG}}-\mu_d\rvert\) for \(\Delta_{\mathrm{sp}}\) along \(\mathcal S_{\mathrm F}\) and \({\bar\Delta}_{\mathrm{sp}}\) along \(\mathcal S_{\mathrm{PG}}\) (Fig.~\ref{fig:2}c). The nodal-antinodal dichotomy is a defining hallmark of the pseudogap.\cite{ref17}

The single-particle Green' s function has an unusual pole-zero structure in the PG regime: at \(T=0\,\mathrm K\), \(G_{\sigma\sigma}^{R}(\mathbf q,\omega)\) has no zero-frequency poles except at the PG node \(\mathbf q_{\mathrm n}\), whereas its zero-frequency zeros lie on \(\xi_{\mathbf q}=\mathcal J_{\mathbf q}/2\), defining \(\mathcal S_{\mathrm L}\), a Luttinger surface.\cite{ref58} In the present model, \(\mathcal S_{\mathrm L}\) is punctured at \(\mathbf q_{\mathrm n}\). These zeros originate from destructive interference between equal-weight electron-removal and electron-addition poles. The finite-temperature spectral function \(A_\Gamma(\mathbf q,\omega)\) contains all four Lehmann branches \(\omega_\lambda=\xi_{\mathbf q}+\lambda \mathcal J_{\mathbf q}\), with \(\lambda=-2,-1,0,+1\), where \(\Gamma\) denotes the energy resolution. Near the PG nodes, the \(\lambda=-1,0\) branches carry the largest weight, and finite broadening produces the two dominant ridges centred near their exact zero-energy contours, \(\xi_{\mathbf q}=\mathcal J_{\mathbf q}\) and \(\xi_{\mathbf q}=0\), respectively, whereas the \(\lambda=-2,+1\) branches produce weaker nearby satellites (Fig.~\ref{fig:2}d). Thus, the visible Fermi arc in \(A_\Gamma(\mathbf q,0)\) represents thermally redistributed spectral weight broadened by finite energy resolution, rather than a continuous locus of genuinely gapless quasiparticle poles.

The temperature and doping evolutions of \(A_\Gamma(\mathbf q,0)\) are displayed in Fig.~\ref{fig:2}e-j. Warming from 50 to 500 K at fixed \(x=0.05\) extends the visible Fermi arc tangentially about each PG node and progressively reveals the subsidiary ridges, but leaves its transverse width nearly unchanged (Fig.~\ref{fig:2}e-g). Increasing \(x\) from 0.05 to 0.20 at fixed \(T=200\,\mathrm K\) makes the visible arc longer and narrower (Fig.~\ref{fig:2}f,h--j). The nodal expansion can estimate this evolution quantitatively. Near \(\mathbf q_{\mathrm n}\), let \(q_\parallel\) and \(q_\perp\) denote displacements tangential and normal to the PG Fermi contour, respectively; then \(\xi\simeq v_{F,\mathrm n}q_\perp\) and \(\mathcal J\simeq\kappa_{\mathrm n}q_\parallel^2\). Within the nodal expansion, the \(\Gamma\)-broadened local line shape collapses onto the dimensionless function \(\mathscr A(v_{\mathrm n},u_{\mathrm n};t_{\mathrm n})=A_\Gamma(\mathbf q,0;T)/A_\Gamma(\mathbf q_{\mathrm n},0;T)\), where \(v_{\mathrm n}=v_{F,\mathrm n}q_\perp/\Gamma\), \(u_{\mathrm n}=\kappa_{\mathrm n}q_\parallel^2/\Gamma\), and \(t_{\mathrm n}=T/\Gamma\). Define the visible Fermi arc as \(\mathscr A\ge1/2\); its half-width and half-length obey \(\lvert q_\perp\rvert_{\mathrm{arc}}=\Gamma/v_{F,\mathrm n}\) and \(\lvert q_\parallel\rvert_{\mathrm{arc}}=\sqrt{\Gamma/\kappa_{\mathrm n}}\,Q_{1/2}(t_{\mathrm n})\), respectively, where \(Q_{1/2}(t_{\mathrm n})\) is a dimensionless crossover function. Thus the arc width is entirely resolution controlled, a consequence of the independent-block structure of \(\hat H_{\mathrm{PG}}\). The arc length scales as \(\sqrt{\Gamma/\kappa_{\mathrm n}}\) in both \(t_{\mathrm n}\ll1\) and \(t_{\mathrm n}\gg1\) limits, with different dimensionless prefactors. However, the absolute arc length remains non-universal, as it depends on the chosen threshold for \(\mathscr A\) and the energy resolution \(\Gamma\).

\protect\phantomsection\label{cooper-pair-surface-lifshitz}
\section{Cooper-surface Lifshitz transition}\label{cooper-surface-lifshitz-transition}

The two-particle Cooper-pair spectrum can nevertheless support gapless excitations along two finite contours in relative-momentum space. In analogy to a conventional Fermi surface, \(\Delta_{\mathrm{pair}}=0\) defines two \emph{Cooper surfaces}, \(\mathcal S_{2e}:\xi_{\mathbf q}=0\) and \(\mathcal S_{2h}:\xi_{\mathbf q}=\mathcal J_{\mathbf q}\), which are level-crossing contours rather than condensates, with \(\mathcal S_{2e}\) coinciding with the PG Fermi contour.
\protect\phantomsection\label{eq-main-6}
\begin{equation}
\label{eq:6}
\Delta_{\mathrm{pair}}(\mathbf q)=
        \begin{cases}
        2\xi_{\mathbf q}-2\mathcal J_{\mathbf q},
        &N_{\mathbf q}=0,\quad \xi_{\mathbf q}>\mathcal J_{\mathbf q},\\
        \min(2\mathcal J_{\mathbf q}-2\xi_{\mathbf q},2\xi_{\mathbf q}),
        &N_{\mathbf q}=2,\quad 0< \xi_{\mathbf q}< \mathcal J_{\mathbf q},\\
        -2\xi_{\mathbf q},
        &N_{\mathbf q}=4,\quad \xi_{\mathbf q}<0,
        \end{cases}
\end{equation}
The two Cooper surfaces are the loci of the zero-frequency poles of the retarded Cooper-pair Green' s function \(G_{\Delta}^{R}(\mathbf q,\Omega)\) at zero temperature. At finite temperature, however, \(\mathcal S_{2e}\) and \(\mathcal S_{2h}\) are more directly resolved in the non-negative pair-removal correlation function \(S_{\Delta}^{<}\), because the Kubo--Martin--Schwinger relation \(A_{\Delta}(\mathbf q,\Omega)=(e^{\beta\Omega}-1)S_{\Delta}^{<}(\mathbf q,\Omega)\) forces the zero-frequency pair spectral function \(A_{\Delta}(\mathbf q,0)\) to vanish wherever \(S_{\Delta}^{<}\) is regular.

\begin{figure}[p]
\centering
\includegraphics[width=0.77\textwidth]{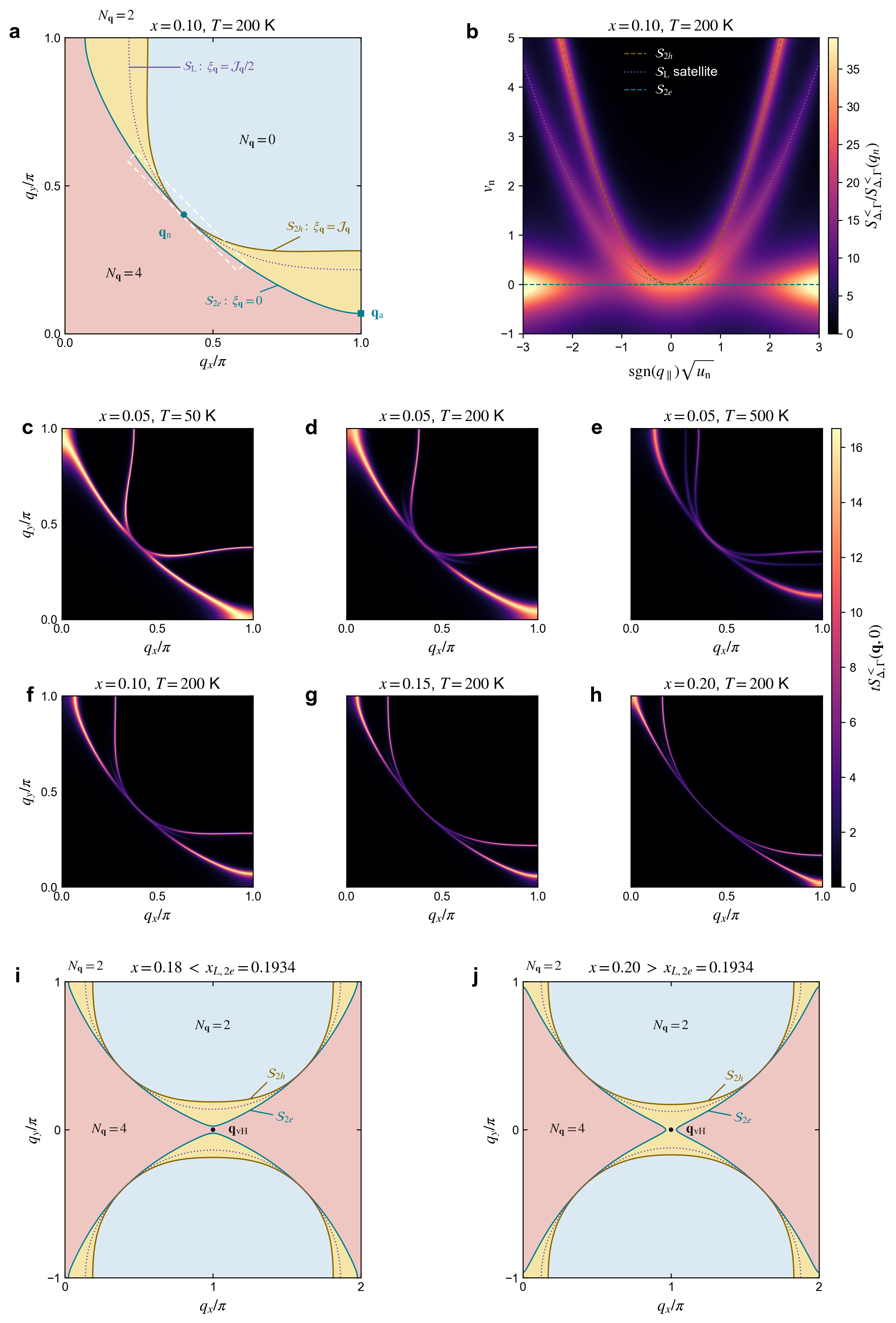}
\captionsetup{width=0.77\textwidth,font=footnotesize,skip=4pt}
\caption{\textbf{Cooper surfaces and Lifshitz transition.} \textbf{a,} Ground-state occupations of \(\hat H_{\mathrm{PG}}\) at \(x=0.10\), annotated with \(N_{\mathbf q}=4,2,0\), the PG node \(\mathbf q_{\mathrm n}\), the antinode \(\mathbf q_{\mathrm a}\), the occupation boundaries \(\mathcal S_{2e}:\xi_{\mathbf q}=0\) and \(\mathcal S_{2h}:\xi_{\mathbf q}=\mathcal J_{\mathbf q}\), and the punctured Luttinger surface \(\mathcal S_{\mathrm L}:\xi_{\mathbf q}=\mathcal J_{\mathbf q}/2\); the white dashed window corresponds to panel b. \textbf{b,} Node-normalized zero-frequency \(S_{\Delta,\Gamma}^{<}(\mathbf q,0)\) at \(x=0.10\) and \(T=200\,\mathrm K\). The \(\mathcal S_{2e}\) and \(\mathcal S_{2h}\) guides are dashed, while the \(\mathcal S_{\mathrm L}\) satellite is dotted; the local coordinates are \(u_{\mathrm n}=\kappa_{\mathrm n}q_\parallel^2/\Gamma\) and \(v_{\mathrm n}=v_{F,\mathrm n}q_\perp/\Gamma\). The two dominant pair-removal ridges follow \(\mathcal S_{2e}\) and \(\mathcal S_{2h}\), whereas the weaker thermally activated intermediate satellite follows \(\mathcal S_{\mathrm L}\). \textbf{c-e,} Non-negative, Lorentzian-broadened zero-frequency pair-removal intensity at fixed \(x=0.05\) for \(T=50,200,500\,\mathrm K\). \textbf{f-h,} Its doping evolution at \(T=200\,\mathrm K\) for \(x=0.10,0.15,0.20\). \textbf{i,j,} Extended-zone contours at \(x=0.18,0.20\). The \(\mathcal S_{2e}\) contour undergoes a neck reconnection as it crosses the van Hove saddle point at \(x_{L,2e}=0.1934\). The displayed spectral maps use \(480\times480\) sampling. The calculations use \(t=0.4\,\mathrm{eV}\), \(t'/t=-0.22\), \(t''/t=0.045\), \(J/t=0.16\), and \(\Gamma/t=0.01\;(4.0\,\mathrm{meV})\).}
\label{fig:3}
\end{figure}

The temperature and doping evolutions of \(S_{\Delta,\Gamma}^{<}(\mathbf q,0;T)\) are displayed in Fig.~\ref{fig:3}c-h. Warming from 50 to 500 K at fixed \(x=0.05\) redistributes weight between the branches associated with the two Cooper surfaces and visibly reveals the middle satellite at \(200\,\mathrm K\): the thermally activated \(N_{\mathbf q}=3\to1\) pair-removal branch, whose zero-frequency contour coincides with \(\mathcal S_{\mathrm L}\) (Fig.~\ref{fig:3}c-e). At fixed \(T=200\,\mathrm K\), increasing doping from \(x=0.10\) to 0.20 weakens \(\mathcal J_{\mathbf q}(x,T)\) and brings \(\mathcal S_{2e}\) and \(\mathcal S_{2h}\) closer together (Fig.~\ref{fig:3}f-h). The nodal expansion quantifies these changes. Near \(\mathbf q_{\mathrm n}\), \(\mathcal S_{2e}\), the \(\mathcal S_{\mathrm L}\) satellite, and \(\mathcal S_{2h}\) approach \(q_\perp=0\), \(q_\perp=\kappa_{\mathrm n}q_\parallel^2/(2v_{F,\mathrm n})\) and \(q_\perp=\kappa_{\mathrm n}q_\parallel^2/v_{F,\mathrm n}\), respectively (Fig.~\ref{fig:3}b), and the \(\Gamma\)-broadened zero-frequency pair-removal intensity collapses onto the node-normalized scaling form \(\mathscr P(v_{\mathrm n},u_{\mathrm n};t_{\mathrm n})=S_{\Delta,\Gamma}^{<}(\mathbf q,0;T)/S_{\Delta,\Gamma}^{<}(\mathbf q_{\mathrm n},0;T)\). At fixed instantaneous \(\xi_{\mathbf q}\) and \(\mathcal J_{\mathbf q}\), temperature redistributes spectral weights through \(t_{\mathrm n}\) without shifting \(\mathcal S_{2e}\), the \(\mathcal S_{\mathrm L}\) satellite, or \(\mathcal S_{2h}\), whereas their normal separations are controlled by \(\kappa_{\mathrm n}/v_{F,\mathrm n}\). In the fixed-density temperature sequence of Fig.~\ref{fig:3}c-e, however, the recalculated \(\mu_{\mathrm{PG}}(T)\) and \(\Phi_d(T)\) can also move these contours in absolute momentum.

The Cooper surfaces undergo a Lifshitz transition by the same topological mechanism as a Fermi surface, despite their different microscopic origin. Let \(\Omega_\zeta(\mathbf q)\) denote the pair dispersion measured relative to the Cooper surfaces \(\mathcal S_\zeta\), with \(\Omega_{2e}=2\xi_{\mathbf q}\) and \(\Omega_{2h}=2(\xi_{\mathbf q}-\mathcal J_{\mathbf q})\). Near a stationary momentum \(\mathbf q_L\), it expands as \(\Omega_\zeta(\mathbf q_L+\delta\mathbf q)=r_\zeta+\frac12\delta\mathbf q^{\mathsf T}\mathcal H_\zeta\delta\mathbf q+O(|\delta\mathbf q|^3)\), where \(r_\zeta=\Omega_\zeta(\mathbf q_L)\), \(\boldsymbol\nabla_{\mathbf q}\Omega_\zeta(\mathbf q_L)=0\) and \((\mathcal H_\zeta)_{ij}=\left.\partial_{q_i}\partial_{q_j}\Omega_\zeta\right|_{\mathbf q_L}\). When \(r_\zeta=0\), \(\mathcal S_\zeta\) passes through \(\mathbf q_L\); provided \(\det\mathcal H_\zeta\ne0\), a sign change of \(r_\zeta\) changes its local topology. In 2D, the sign of \(\det\mathcal H_\zeta\) determines the transition type: \(\det\mathcal H_\zeta<0\) gives a saddle-point neck reconnection, whereas \(\det\mathcal H_\zeta>0\) gives the creation or annihilation of a closed pocket at a local extremum. At the nondegenerate van Hove saddle point \(\mathbf q_{\mathrm{vH}}=X,Y\), square-lattice symmetry enforces \(\boldsymbol\nabla_{\mathbf q}\Omega_\zeta(\mathbf q_{\mathrm{vH}})=0\). Thus \(\mathbf q_{\mathrm{vH}}\) is a symmetry-enforced stationary point rather than an accidental stationary point produced by tuning. \(\mathcal S_\zeta\) passes through the stationary point \(\mathbf q_{\mathrm{vH}}\) when \(\Omega_\zeta(\mathbf q_{\mathrm{vH}})=0\), and this crossing through the van Hove saddle point corresponds to a neck-reconnection Lifshitz transition. For the parameters used here, the \(\mathcal S_{2e}\) neck reconnection occurs at \(x_{L,2e}=0.1934\) (Fig.~\ref{fig:3}i,j); no \(\mathcal S_{2h}\) crossing is found at the seven sampled dopings spanning \(0.01\le x\le0.30\). The Lifshitz critical doping \(x_{L,\zeta}\) is rather sensitive to the parameters. Note that the Lifshitz transition in relative-momentum space does not by itself generate a condensate amplitude or phase stiffness.

The exact block solution permits thermodynamic quantities of \(\hat H_{\mathrm{PG}}\) to be evaluated exactly. At low temperatures, the dominant low-energy contributions arise from Cooper surface and nodal patches. Away from the Lifshitz transition, regular Cooper surface patches give \(c_V^{\mathrm{pair}}/T,\kappa_c^{\mathrm{pair}}\propto\rho_{\mathrm{pair}}(0)\), where \(\rho_{\mathrm{pair}}(E)\) is the density of states (DOS) near the Cooper surfaces, but make no leading contribution to the spin response as Cooper-pair excitations are spin-singlets. Nodal patches have DOS \(\rho_{\mathrm{nodal}}(E)\propto|E|^{1/2}\), so \(c_V^{\mathrm{nodal}}/T,\chi_s^{\mathrm{nodal}},\kappa_c^{\mathrm{nodal}}\propto T^{1/2}\). Consequently, the spin Wilson ratio \(R_W^s\propto\chi_s/(c_V/T)\propto T^{1/2}\to0\). Near the Lifshitz critical doping, the singular contributions are \(c_{V,\mathrm s}/T,\kappa_{c,\mathrm s}\propto\ln[E_{\mathrm{cut}}/\max(|r_{2e}|,T)]\), where \(E_{\mathrm{cut}}\) is the ultraviolet cutoff, while the spin susceptibility has no singular contribution. Thus \(R_W^s(x_{L,2e})\propto T^{1/2}/\ln(E_{\mathrm{cut}}/T)\to0\). Taking \(r_{2e}\) as the independent detuning, the singular grand-potential density obeys \(f_{\mathrm s}(r_{2e},0)\propto-r_{2e}^2\ln(E_{\mathrm{cut}}/|r_{2e}|)\) and \(f_{\mathrm s}(0,T)\propto-T^2\ln(E_{\mathrm{cut}}/T)\), giving the Lifshitz exponent \(\alpha_{\mathrm L}=0\) with a multiplicative logarithm, which arises directly from the 2D saddle-point DOS. The quadratic saddle point defines the kinematic energy-momentum scaling exponents \(z_{\mathrm L}=2\) and \(\nu_{\mathrm L}=1/2\), which at leading powers obey the relation \(2-\alpha_{\mathrm L}=\nu_{\mathrm L}(d+z_{\mathrm L})=2\) for \(d=2\), formally analogous to quantum hyperscaling. However, \(\nu_{\mathrm L}\) is a relative-momentum detuning exponent rather than a real-space correlation-length exponent, \(z_{\mathrm L}\) does not describe the dynamics of a propagating collective mode, and no symmetry is broken.

\protect\phantomsection\label{ohmic-mfl}
\section{Ohmic dissipation and MFL}\label{ohmic-dissipation-and-mfl}

The dynamic SM organization is the complementary finite-\((\mathbf Q,\Omega)\) organization from the same Cooper channel of AFM superexchange and is characterized physically by a broad retarded pair spectrum \(\mathcal A_{\alpha\alpha}^{\mathrm{pair}}(\mathbf Q,\Omega)\). An exact Hubbard--Stratonovich transformation of \(-2g_J(x)JN_s\sum_{\mathbf Q,\alpha}\hat\Lambda_{\mathbf Q,\alpha}^{\dagger}\hat\Lambda_{\mathbf Q,\alpha}\) introduces an auxiliary pair field \(\bar\Xi_{\mathbf Q,\alpha}(\tau)\) for each \((\mathbf Q,\alpha)\) sector. Integrating out the fermionic fields and expanding the fermion determinant to quadratic order gives the inverse propagator of \(\bar\Xi\), \([D_{\bar\Xi}^{-1}(\mathbf Q,i\Omega_m)]_{\alpha\alpha'}=\delta_{\alpha\alpha'}/[2g_J(x)J]-\Pi_{\alpha\alpha'}(\mathbf Q,i\Omega_m)\). Here \(\Pi_{\alpha\alpha'}\) is the particle-particle polarization. In the Gaussian approximation, self-energy feedback and vertex corrections are not included, and the low-energy behaviour is governed mainly by the infrared kinematics near the Fermi surface. After analytic continuation, the Ohmic dissipation coefficient is \(\gamma_{\alpha\alpha'}(\mathbf Q,T)=\lim_{\Omega\to0^+}\operatorname{Im}\Pi_{\alpha\alpha'}^R(\mathbf Q,\Omega)/\Omega\). For \(\bar v_F|\mathbf Q|\ll T\), the particle-particle polarization approaches the low-energy continuum \(\lim_{\mathbf Q\to\mathbf 0}\operatorname{Im}\Pi_{\alpha\alpha'}^R(\mathbf Q,\Omega)\propto\tanh(\Omega/4T)\), yielding \(\gamma_{\alpha\alpha'}(\mathbf Q,T)\allowbreak\propto T^{-1}\), where \(\bar v_F\) is a characteristic renormalized Fermi-velocity scale. For \(T\ll\bar v_F|\mathbf Q|\ll E_F\), two \(\mathbf Q\)-shifted renormalized Fermi surfaces instead intersect transversely, generating an approximately temperature-independent Ohmic dissipation coefficient. The Ohmic damping here is generated by a particle-particle bubble and is distinct from the particle-hole-bubble-generated damping in the Hertz--Millis theory of itinerant metallic quantum criticality.\cite{ref59,ref60}

Finite-\(\mathbf Q\), approximately temperature-independent Ohmic damping provides the dissipative mechanism for the SM. The physical picture is that finite-COM collective Cooper-pair fluctuations decay into the gapless particle-particle continuum. A complementary dynamic RML prescription provides the block-local pair field \(\Xi_{\mathbf q}\) with inverse propagator \(D_{\Xi,0}^{-1}(i\Omega_m)=r_s+i\ell_s\Omega_m+\gamma_s|\Omega_m|\), which is postulated to have the same infrared form as \(D_{\bar\Xi}^{-1}\) for \(|\Omega_m|\ll E_F\). \(\Xi_{\mathbf q}\) may microscopically arise from a weighted average of the collective pair kernel over both extended-\(s\)- and \(d\)-wave channels and over an annulus of finite but small \(|\mathbf Q|\) satisfying the transverse-intersection condition, thereby yielding a channel-independent \(\Xi_{\mathbf q}\). In addition to the mode reorganization, the dynamic RML prescription also supplies the isotropic fermion-pair coupling \(S_{c\Xi}=-2J_s\int_0^\beta d\tau\sum_{\mathbf q}'[\Xi_{\mathbf q}^{*}(\tau)\Delta_{\mathbf q}(\tau)+\mathrm{h.c.}]\), where \(\Delta_{\mathbf q}(\tau)\) is the path-integral coherent-state representation of \(\hat\Delta_{\mathbf q}\). Integrating out the bosonic field \(\Xi_{\mathbf q}\) from the SM effective action \(S_s=S_c+S_\Xi+S_{c\Xi}\) yields the leading infrared one-loop fermion decay rate
\protect\phantomsection\label{eq-main-7}
\begin{equation}
\label{eq:7}
-\operatorname{Im}\Sigma_s^R(\mathbf q_F,\omega)
        \simeq\lambda_s\omega\coth\!\left(\frac{\omega}{2T}\right)
        \longrightarrow
        \begin{cases}
        2\lambda_sT,&|\omega|\ll T\ll\Omega_{\mathrm{dyn}},\\
        \lambda_s|\omega|,&T\ll|\omega|\ll\Omega_{\mathrm{dyn}},
        \end{cases}
\end{equation}
where \(\Sigma_s^R\) is the retarded one-loop fermionic self-energy, \(\mathbf q_F\) lies on the renormalized Fermi surface, \(\lambda_s=J_s^2\gamma_s/r_s^2\), and \(\Omega_{\mathrm{dyn}}=r_s/\sqrt{\ell_s^2+\gamma_s^2}\). The decay rate is linear in \(T\) for \(|\omega|\ll T\) and linear in \(|\omega|\) for \(T\ll|\omega|\), exhibiting the characteristic single-particle scaling of an MFL self-energy in the SM.\cite{ref52}

In the quantum regime \(T\ll|\omega|\ll\Omega_{\mathrm{dyn}}\), applying the Kramers--Kronig relation to \(\operatorname{Im}\Sigma_s^R\) yields the leading logarithmic contribution, \(\operatorname{Re}\Sigma_s^R(\mathbf q_F,\omega)\sim-(2\lambda_s/\pi)\omega\ln(\Omega_{\mathrm{UV}}/|\omega|)\), where \(\Omega_{\mathrm{UV}}\) is set by the causal ultraviolet completion. The quasiparticle weight \(Z_s(\omega)=[1-\partial_\omega\operatorname{Re}\Sigma_s^R]^{-1}\sim[1+(2\lambda_s/\pi)\ln(\Omega_{\mathrm{UV}}/|\omega|)]^{-1}\) therefore vanishes logarithmically in the limit \(|\omega|\to0\) subject to \(T/|\omega|\to0\). The vanishing \(Z_s(\omega)\), together with \(-\operatorname{Im}\Sigma_s^R\sim\lambda_s|\omega|\), which remains comparable to the single-particle excitation energy \(|\omega|\), precludes asymptotically sharp Landau quasiparticle poles and renders the spectral response near the renormalized Fermi surface incoherent. In the thermal regime \(|\omega|\ll T\ll\Omega_{\mathrm{dyn}}\), temperature instead cuts off the logarithm, \(\operatorname{Re}\Sigma_s^R(\mathbf q_F,\omega,T)\sim-(2\lambda_s/\pi)\omega\ln(\Omega_{\mathrm{UV}}/T)\), leading to a finite \(Z_s(T)\). Nevertheless, the single-particle spectral function retains a finite thermal width \(-\operatorname{Im}\Sigma_s^R\simeq2\lambda_sT\) that is much larger than \(|\omega|\). A finite \(Z_s(T)\) therefore does not imply a long-lived Landau quasiparticle, and the spectral response remains incoherent. Across both regimes, the renormalized Fermi surface remains a momentum-space locus of enhanced low-energy spectral weight, but no longer admits a Landau-quasiparticle description.

\protect\phantomsection\label{crossover-phase-transitions}
\section{Crossover and phase transitions}\label{crossover-and-phase-transitions}

The PG-SM crossover is governed by spectral-weight transfer (SWT) within the physical pair spectrum \(\mathcal A_{\alpha\alpha}^{\mathrm{pair}}(\mathbf Q,\Omega)\). With increasing \(x\), the SC amplitude \(\Phi_d(x)\) decreases, whereas the Gutzwiller factor \(g_t(x)\) increases. These opposing doping trends correspondingly weaken the quasi-static PG component while increasing the relevance of the Ohmic damping from the finite-\(\mathbf Q\) particle-particle continuum in the SM, thereby providing a qualitative microscopic rationale for the spectral-weight redistribution. The static and dynamic RML prescriptions encode two limiting organizations in the PG Hamiltonian \(\hat H_{\mathrm{PG}}\) and the SM effective action \(S_s\), respectively, but neither side has global phase coherence, and the crossover breaks no symmetry. A proposed window-dependent diagnostic of the PG-SM crossover is the SWT fraction \(f_{\mathrm{SWT}}(x,T)=W_{\mathrm{SM}}/(W_{\mathrm{PG}}+W_{\mathrm{SM}})\), which increases as weight moves from the PG to the SM. \(W_{\mathrm{PG}}\) and \(W_{\mathrm{SM}}\) are obtained by integrating the same positive-frequency spectrum over the common window \(0<\Omega<\Omega_c\) and the fixed COM regions \(|\mathbf Q|< Q_0\) and \(Q_1<|\mathbf Q|< Q_2\), respectively, with \(Q_0< Q_1< Q_2\). The maximum of \(|\partial_T f_{\mathrm{SWT}}|\) at fixed \(x\) can define an operational \(T^*(x)\), with \(|\partial_x f_{\mathrm{SWT}}|\) providing the analogous criterion in a doping scan. Deriving both RML prescriptions from a common dressed Cooper kernel of the \(t\)-\(J\) model remains an open problem.

The PG-SC transition begins by restoring the off-diagonal scattering between distinct blocks through \(\hat H_{\mathrm{PG}}^{\mathrm{off}}=-\sum_{\mathbf q\ne\mathbf q'}'\mathcal K_{\mathbf q\mathbf q',\alpha}(x)\hat\Delta_{\mathbf q}^{\dagger}\hat\Delta_{\mathbf q'}\), where a separable coupling with \(\alpha\)-channel symmetry is \(\mathcal K_{\mathbf q\mathbf q',\alpha}(x)=\Phi_\alpha(x)\phi_{\mathbf q,\alpha}\phi_{\mathbf q',\alpha}/N_s\). A mean-field treatment of \(\hat H_{\mathrm{PG}}^{\mathrm{off}}\), solved self-consistently together with \(\hat H_{\mathrm{PG}}\), determines an anomalous saddle at which inter-block coherence is restored. The coherent blocks then superpose into the collective mode \(\Phi_\alpha=2g_J(x)J N_s^{-1}\sum_{\mathbf q}\phi_{\mathbf q,\alpha}\langle\hat\Delta_{\mathbf q}\rangle\), and the saddle acquires a nonzero bare phase stiffness, thereby recovering the conventional BCS description. The distinguishing ingredient is that the coherent saddle develops on top of the PG pair correlations encoded in the diagonal PG spectral weight. In 2D, the collective mode acquires global phase coherence and enters the SC state only below \(T_{\mathrm{BKT}}\).

The SM-SC transition follows an analogous route in the path-integral formalism: an off-diagonal field term \(S_\Xi^{\mathrm{off}}\) couples the block-local fields, and retaining the quartic term gives a Ginzburg--Landau expansion whose quadratic coefficient \(r_{\mathrm{SC}}\) determines whether a nonzero saddle exists. Because the SM inverse propagator contains the Ohmic term \(\gamma_s|\Omega_m|\), the resulting time-dependent Ginzburg--Landau action describes overdamped relaxational dynamics, whose critical \(r_{\mathrm{SC}}=0\) limit is diffusive. For \(r_{\mathrm{SC}}<0\) and a positive quartic coefficient, the block-local fields coherently superpose into a collective field. Expanding about this saddle reduces the block-local theory to a conventional Ginzburg--Landau action for the collective field and yields a nonzero bare phase stiffness. In 2D, vortex binding at \(T_{\mathrm{BKT}}\) establishes algebraic phase coherence and a nonzero renormalized stiffness, thereby completing the transition to the SC state.

\section{Discussion}\label{discussion}

\subsection{Theoretical aspects}\label{theoretical-aspects}

The central microscopic task is to derive both RML prescriptions. The PG side exposes a specific matching ambiguity: a physical quasi-static collective peak has finite widths \(\Delta Q\sim\xi_{\mathrm{PG}}^{-1}\) and \(\Delta\Omega\sim\tau_{\mathrm{PG}}^{-1}\), whereas \(\hat H_{\mathrm{PG}}\) retains strictly zero-COM pair operators. A normalized weighted average over \(|\mathbf Q|\lesssim\Delta Q\) and \(|\Omega|\lesssim\Delta\Omega\) is a possible matching prescription, but the reduction to \(\mathcal J_{\mathbf q}\delta_{\mathbf q\mathbf q'}\) remains to be derived. Neither BKT phase disordering, the Cooper logarithm nor the \(\mathbf Q\to0\) limit establishes relative-momentum locality. \(\hat H_{\mathrm{PG}}\) therefore does not determine finite-\(\mathbf Q\) pair propagation and lifetime. The dynamic RML prescription likewise requires a finite-\(\mathbf Q\) sector with transverse intersections, a positive-definite pair-mass matrix and an approximately isotropic and temperature-independent Ohmic coefficient. Reducing the finite-\(\mathbf Q\) sector to a channel-independent \(\Xi_{\mathbf q}\) requires microscopic justification, while the MFL benchmark further requires \(r_s>0\), a causal ultraviolet completion, a non-collapsing scaling window \(\max(T,|\omega|)\ll\Omega_{\mathrm{dyn}}\), and an approximately temperature-independent \(\lambda_s\).

The particle-particle bubble in the Gaussian approximation and the one-loop self-energy do not constitute a closed many-body theory. Under feedback, a bubble already embedded in the matched kernel must be replaced rather than added again; any subtraction or counterterm must be introduced in the same stationary functional. A large-scale self-consistent calculation formulated within a Baym--Kadanoff-conserving framework, using a two-particle-irreducible \(GG\) \(T\)-matrix, can provide a minimal benchmark for fermionic self-energy feedback and consistent pair vertices. When \(\lambda_s\ln(\Omega_{\mathrm{UV}}/|\omega|)\) reaches \(O(1)\), higher-order logarithmic corrections become comparable to the one-loop term, so the perturbative expansion loses control and the present MFL result may be modified. A gauge-invariant theory of incoherent transport in the SM, incorporating momentum relaxation and response vertices related to the fermionic self-energy by the Ward identity, remains to be developed to establish whether the linear-\(T\) single-particle decay rate produces \(T\)-linear dc transport.\cite{ref28}

In the spirit of Landau Fermi-liquid theory, an analogous \emph{Cooper liquid} organized around the Cooper surfaces of the exactly solvable \(\hat H_{\mathrm{PG}}\) is conjectured. Generic inter-block interactions remove block factorization and hence exact solvability; however, if they can be introduced adiabatically without an intervening phase transition while the zero-frequency pair poles remain sharp, the elementary low-energy excitations of the interacting system, here termed \emph{quasi-pairs}, could remain adiabatically connected to the exact block-pair states of \(\hat H_{\mathrm{PG}}\) near the Cooper surfaces. Establishing this continuity and determining the quasi-pair residue, lifetime and residual interactions remain open problems. The corresponding central issue is the instability of the Cooper surfaces. A concrete mean-field treatment of \(\hat H_{\mathrm{PG}}^{\mathrm{off}}\) shows that a nonzero anomalous saddle can open a pair gap on a regular Cooper surface; for a \(d\)-wave saddle, symmetry-protected nodes remain. Note that the Cooper-surface gap lies in relative-momentum space.

\subsection{Experimental aspects}\label{experimental-aspects}

Pair-sensitive spectroscopy can directly probe the Cooper surfaces. The sharpest spectroscopic signature is a zero-energy charge-\(2e\) ridge, accessible in principle to 2e-ARPES\cite{ref61,ref62}, without a corresponding off-node charge-\(e\) pole in the PG regime. A momentum-resolved pair-removal probe that couples to \(\hat\Delta_{\mathbf q}\) should resolve broadened ridges on \(\mathcal S_{2e}\) and \(\mathcal S_{2h}\), which meet at the PG nodes, whereas the zero-temperature normal-state limit has no off-node electron pole; finite-temperature ARPES spectra may nevertheless show thermally activated or resolution-broadened weight, appearing as the so-called Fermi arc. The absence of the ridges would constitute a falsification of the Cooper-surface prediction only if the relevant pair transitions are symmetry allowed, their matrix elements are appreciable and the momentum--energy resolution is sufficient.

Identifying the Cooper-surface Lifshitz transition requires combined spectroscopic and thermodynamic measurements. The phase-incoherent normal-state limit predicts finite-temperature-rounded logarithmic enhancements of \(c_V/T\) and the charge response near the Lifshitz critical doping, but no logarithmic enhancement in either the spin susceptibility or the spin contribution to the Knight shift. SC will generally pre-empt the \(T\to0\) limit, so accessing the Lifshitz critical regime requires suppressing coherence, with the caveat that the applied field or other tuning may reconstruct the spectrum. Coincident observations of pair-ridge reconnection and charge-spin asymmetry would support a Cooper-surface Lifshitz transition interpretation. The numerical \(x_{L,2e}\) is parameter sensitive and should not be identified with the PG critical doping.

In the SM, the Gaussian kinematic benchmark has two COM-momentum regimes: \(\gamma_{\alpha\alpha'}(\mathbf Q,T)\allowbreak\propto T^{-1}\) for \(\bar v_F|\mathbf Q|\ll T\), but becomes approximately temperature independent for \(T\ll\bar v_F|\mathbf Q|\ll E_F\). A decisive test should observe the dressed finite-COM pair continuum and the associated single-particle MFL response in the same doping and temperature regime. The dynamic RML prescription predicts an \(\omega/T\) collapse of \(-\operatorname{Im}\Sigma_s^R/T\), accompanied by the Kramers--Kronig logarithm in \(\operatorname{Re}\Sigma_s^R\), an approximately isotropic self-energy coefficient \(\lambda_s\) along the renormalized Fermi surface, and a finite scaling window as temperature decreases. The operational \(f_{\mathrm{SWT}}\), extracted from fixed \((\mathbf Q,\Omega)\) windows, depends on the chosen windows and is not a sum rule; it quantifies the growth of spectral weight in the dressed finite-COM pair continuum, while maxima of \(|\partial_Tf_{\mathrm{SWT}}|\) or \(|\partial_xf_{\mathrm{SWT}}|\) define operational crossover loci. SC must instead be identified by nonzero phase stiffness and global phase coherence.

\section*{Acknowledgements}
J.-J. M. thanks Cui Ding for helpful discussions and Tao Li for sharing lecture slides from his course on high-temperature superconductivity. This work was supported by the National Natural Science Foundation of China (Grant No. 12404171) and the Guangdong Project (Grant No. 2024QN11X176).

\providecommand{\doibase}{https://doi.org/}
\makeatletter
\immediate\write\@auxout{\string\citation{apsrev41Control}}
\makeatother
\bibliographystyle{apsrev4-1}
\bibliography{references}

\end{document}